\documentclass[12pt]{iopart}
\usepackage{iopams,epsfig}

\newcommand{\APJ}{{\it Astrophys.~J.\ }}
\newcommand{\APJS}{{\it Astrophys.~J.\ Suppl.\ }}
\newcommand{\AAP}{{\it Astron.\ Astrophys.\ }}

\newcommand{\astroph}[1]{{\it Preprint} astro-ph/#1} 
\newcommand{\arXiv}[1]{{\it Preprint} arXiv:#1} 

\newcommand{\PF}{{\it Phys.\ Fluids\ }}
\newcommand{\PFB}{{\it Phys.\ Fluids~B\ }}
\newcommand{\PP}{{\it Phys.\ Plasmas\ }}
\newcommand{\JGR}{{\it J.~Geophys.\ Res.\ }}

\newcommand{\eqref}[1]{\eref{#1}}

\newcommand{\eqsdash}[2]{\eref{#1}--\eref{#2}}
\newcommand{\eqsand}[2]{\eref{#1} and \eref{#2}}
\newcommand{\Eqref}[1]{Equation \eref{#1}}

\newcommand{\figref}[1]{figure \ref{#1}}

\newcommand{\secref}[1]{\S\,\ref{#1}}

\newcommand{\bea}{\begin{eqnarray}}
\newcommand{\eea}{\end{eqnarray}}
\newcommand{\beq}{\begin{equation}}
\newcommand{\eeq}{\end{equation}}
\newcommand{\lt}{\left}
\newcommand{\rt}{\right}
\newcommand{\la}{\langle}
\newcommand{\ra}{\rangle}

\newcommand{\dd}{\partial}
\newcommand{\vdel}{\bnabla}

\newcommand{\ephi}{\varphi}

\newcommand{\vr}{\bi{r}}
\newcommand{\vR}{\bi{R}}
\newcommand{\vk}{\bi{k}}
\newcommand{\vu}{\bi{u}}
\newcommand{\vuperp}{\vu_\perp}
\newcommand{\uperp}{u_\perp}
\newcommand{\upar}{u_\parallel}

\newcommand{\vf}{\bi{f}}
\newcommand{\vE}{\bi{E}}

\newcommand{\vj}{\bi{j}}
\newcommand{\vja}{\vj_\mathrm{ext}}

\newcommand{\vA}{\bi{A}}
\newcommand{\vB}{\bi{B}}
\newcommand{\dvB}{\delta\vB}
\newcommand{\dvBperp}{\delta\vB_\perp}
\newcommand{\dB}{\delta B}
\newcommand{\dBperp}{\dB_\perp}
\newcommand{\dBpar}{\dB_\parallel}

\newcommand{\vz}{\hat{\bi{z}}}

\newcommand{\vdperp}{\vdel_\perp}

\newcommand{\kpar}{k_\parallel}
\newcommand{\kperp}{k_\perp}
\newcommand{\mfp}{\lambda_\mathrm{mfp}}
\renewcommand{\Re}{\mathrm{Re}}

\newcommand{\ul}{\delta u_{\lambda}}
\newcommand{\taul}{\tau_{\lambda}}
\newcommand{\taur}{\tau_{\rho_i}}
\newcommand{\eps}{\varepsilon}

\newcommand{\lparl}{l_{\parallel\lambda}}
\newcommand{\lvisc}{l_\nu}

\newcommand{\urms}{u_\mathrm{rms}}
\newcommand{\lf}{L}
\newcommand{\leps}{l_0}

\newcommand{\kc}{k_{\perp c}}
\newcommand{\dvc}{\delta v_{\perp c}}
\newcommand{\vths}{v_{\mathrm{th}s}}
\newcommand{\vthi}{v_{\mathrm{th}i}}

\newcommand{\vv}{\bi{v}}
\newcommand{\vvperp}{\vv_\perp}
\newcommand{\vperp}{v_\perp}
\newcommand{\vpar}{v_\parallel}
\newcommand{\nui}{\nu_{ii}}
\newcommand{\nuss}{\nu_{ss}}

\newcommand{\qs}{q_s}
\newcommand{\qi}{q_i}
\newcommand{\fMs}{F_{0s}}
\newcommand{\fMi}{F_{0i}}
\newcommand{\fMe}{F_{0e}}
\newcommand{\dfs}{\delta f_s}

\newcommand{\dfe}{\delta f_e}
\newcommand{\hs}{h_s}
\newcommand{\hi}{h_i}
\newcommand{\hik}{h_i(\vk)}
\newcommand{\hil}{h_{i\lambda}}
\newcommand{\phik}{\ephi(\vk)}
\newcommand{\avphik}{\avphi(\vk)}
\newcommand{\phil}{\ephi_\lambda}
\newcommand{\he}{h_e}
 
\newcommand{\avchi}{\la\chi\ra_{\vR_s}}

\newcommand{\avphi}{\la\ephi\ra_{\vR_i}}
\renewcommand{\ns}{n_{0s}}
\renewcommand{\ni}{n_{0i}}

\newcommand{\Ts}{T_{0s}}
\newcommand{\Tsp}{T_{0s'}}
\newcommand{\Ti}{T_{0i}}
\newcommand{\Te}{T_{0e}}

\newcommand{\dtcolls}{\lt({\dd f_s\over\dd t}\rt)_\mathrm{c}}
\newcommand{\dMtcolls}{\lt({\dd\fMs\over\dd t}\rt)_\mathrm{c}}
\newcommand{\ddtcolls}{\lt({\dd\dfs\over\dd t}\rt)_\mathrm{c}}
\newcommand{\dhtcolls}{\lt({\dd\hs\over\dd t}\rt)_\mathrm{c}}
\newcommand{\dhtcolli}{\lt({\dd\hi\over\dd t}\rt)_\mathrm{c}}

\newcommand{\intrV}{\int {\rmd^3\vr\over V}\,}
\newcommand{\intRs}{\int {\rmd^3\vR_s\over V}\,}

\newcommand{\intr}{\int {\rmd^3\vr}\,}
\newcommand{\intv}{\int {\rmd^3\vv}\,}

\begin{document}

\title[Gyrokinetic turbulence: a nonlinear route to dissipation through phase space]{Gyrokinetic turbulence: a nonlinear route to dissipation through phase space}

\author{A~A~Schekochihin,$^1$ S~C~Cowley,$^{1,2}$ W~Dorland,$^3$ 
G~W~Hammett,$^4$ G~G~Howes,$^5$ G~G~Plunk,$^6$ E~Quataert$^5$\\ and T~Tatsuno$^3$}

\address{$^1$ Plasma Physics, Blackett Laboratory, Imperial College, London~SW7~2AZ, UK}
\address{$^2$ UKAEA Culham Division, Abington OX14 3DB, UK}
\address{$^3$ Department of Physics, IREAP and CSCAMM, 
University of Maryland, College Park, MD~20742-3511, USA}
\address{$^4$ Princeton Plasma Physics Laboratory, Princeton, NJ~08543-0451, USA}
\address{$^5$ Department of Astronomy, University of California, Berkeley, CA~94720-3411, USA}
\address{$^6$ Department of Physics and Astronomy, UCLA, Los Angeles, CA~90095-1547, USA}

\ead{a.schekochihin@imperial.ac.uk}

\begin{abstract}
This paper describes a conceptual framework for understanding kinetic plasma 
turbulence as a generalized form of energy cascade in phase space. 
It is emphasized that conversion of turbulent energy into thermodynamic heat 
is only achievable in the presence of some (however small) 
degree of collisionality. The smallness of the collision rate is compensated 
by the emergence of small-scale structure in the velocity space. For gyrokinetic 
turbulence, a nonlinear perpendicular phase mixing mechanism is identified and described as 
a turbulent cascade of entropy fluctuations simultaneously occurring 
at spatial scales smaller than the ion gyroscale and in velocity space. 
Scaling relations for the resulting fluctuation spectra are derived. 
An estimate for the collisional cutoff is provided. The importance of 
adequately modeling and resolving collisions in gyrokinetic simulations is 
biefly discussed, 
as well as the relevance of these results to understanding the dissipation-range 
turbulence in the solar wind and the electrostatic microturbulence in fusion plasmas.
\end{abstract}

\pacs{52.30.Gz, 52.35.Ra, 96.50.Tf\\\\
\PPCF {\bf 50}, 124024 (2008); E-print {\tt arXiv:0806.1069}\\ 
(invited talk for the 35th EPS Conference on Plasma Physics, Crete, 10 June 2008)}

\section{Turbulence: the Symptoms and the Cause}
\label{sec_intro}

What is turbulence? Modulo many definitional and interpretational subtleties \cite{Frisch,Tsinober}, 
{\em turbulence is multiscale disorder}: we tend to say that we are dealing with a turbulent system 
if we have detected (measured, observed, simulated, intuited) chaotic fluctuations of some 
field(s) over a broad range of scales. In plasmas, these fluctuating fields are the
electric and magnetic fields and the distribution function of the particles 
(either measured directly or accessible partially 
via its moments: density, flow velocity, temperature). 
So turbulence is defined as a {\em syndrome} \cite{Stewart_movie}:
it is identifed by its symptoms.\footnote{We thank 
T A Yousef for bringing to our attention this analogy, which is particularly apt in 
fusion contexts, where turbulence is indeed a disease that gives rise to anomalous transport, 
prevents plasma confinement and thus hampers humanity's progress 
toward the hydrogen-powered future.} 
The next logical step is to ask what causes 
the development of the problem in the first place. 
The short answer is energy injection: in  
physical systems, turbulence is stirred up by some source of energy, 
which is system-specific and can be in the form of direct mechanical 
forcing (spoon in a tea cup, supernovae in the 
interstellar medium), boundary conditions (airplane wing), 
or various instabilities feeding on 
background equilibrium gradients (tokamak microturbulence, 
solar convection, magnetorotational turbulence in accretion discs). 
The fluctuation energy injected into the system is nearly always dissipated 
into heat. Because the dissipation mechanisms available to the system 
have to do with its material properties (microphysics) and are usually 
unrelated to the energy-injection mechanism (macrophysics), there is 
more often than not a scale separation between the energy-injection, 
or energy-containing, scale (the {\em outer scale}) and the much smaller dissipation 
scale (the {\em inner scale}). In order to dissipate energy, the system has 
to bridge this gap and one way for this to happen is for the nonlinear interactions   
to fill the intermediate scale range 
with fluctuations --- giving rise to multiscale disorder, or turbulence
(there are, of course, other ways, e.g., shock or current-sheet formation, but we will 
not consider them here). 

The simplest illustration of the argument made above is the case 
of a Navier-Stokes neutral fluid, whose velocity field $\vu$ satisfies 
\bea
\label{NS}
\dd_t\vu + \vu\cdot\vdel\vu = -\vdel p + \nu\nabla^2\vu + \vf,
\quad \vdel\cdot\vu = 0,
\eea
where $p$ is pressure, 
$\nu$ the molecular viscosity of the fluid, and the body force $\vf$ stands in 
for the outer-scale energy injection. The kinetic energy of the fluid then satisfies
\bea
{\rmd\over \rmd t}\intrV {u^2\over2} = \eps - \nu\intrV |\vdel\vu|^2,
\label{NS_energy}
\eea
where $V$ is the system volume and $\eps = (1/V)\intr \vu\cdot\vf$ 
is the injected power per unit volume. In a stationary state, the 
injection and dissipation terms on the right-hand side of this equation 
must balance, even though $\eps$ is finite and viscosity is small,  
or, more precisely, the viscous term in \eqref{NS} is negligible at the outer scale.  
The balance is accomplished by transferring kinetic energy to small scales, 
where the velocity gradients are large, compensating for the viscosity's 
smallness. The viscous (inner) scale to which the energy has to travel 
in order to be dissipated is, on dimensional grounds, 
$\lvisc\sim(\nu^3/\eps)^{1/4}\sim\lf\Re^{-3/4}$, where 
$\lf$ is the outer scale and $\Re=\urms\lf/\nu$ is the Reynolds number. 
The system becomes turbulent when $\Re\gg1$, i.e.,
$\lvisc\ll\lf$, so fluctuations arise over a broad band of 
scales.\footnote{{\em A one-paragraph review of the 
Kolmogorov--Obukhov 1941 turbulence theory} \cite{K41,Obukhov}: 
If it can be assumed (by no means an automatic certainty!) 
that the energy is transported {\em locally} from scale to scale \cite{Richardson}, 
the energy flux through the intermediate scales $\lf\gg\lambda\gg\lvisc$ 
(the {\em inertial range}) must be constant and equal to $\eps$. 
Assuming that fluctuations are isotropic in the inertial range, 
we have $\ul^2/\taul \sim \eps$, where $\ul$ is the characteristic 
relative velocity of fluid elements separated by a distance $\lambda$ and 
$\taul$ is the characteristic nonlinear interaction time ({\em energy-cascade time}) 
at this scale. For a local cascade, dimensionally, $\taul\sim\lambda/\ul$
and the Kolmogorov scaling law immediately follows: $\ul\sim(\eps\lambda)^{1/3}$.
\label{fn_K41}}

\section{Plasma Turbulence: Entropy, Heating and the Kinetic Cascade} 
\label{sec_gen}

Can this argument be generalized to plasma turbulence? 
If the plasma is sufficiently collisional, its dynamics is described 
by a set of fluid equations with diffusive dissipation terms \cite{Braginskii}. 
While things become more complicated than for the Navier--Stokes equation 
(multiple fields and species, different diffusion coefficients perpendicular 
and parallel to the magnetic field, interplay between waves and nonlinear 
interactions), the basic principle remains the same: 
small-scale spatial structure is generated so that the energy injected 
at the outer scale can be transferred to the smaller dissipative scales 
and converted into heat. All this, however, is only valid for fluctuations 
whose characteristic spatial and temporal scales remain collisional, 
namely $\kpar\mfp\ll1$ and $\omega\ll\nui$, where $\kpar$ is 
the typical wavenumber parallel to the magnetic field, $\mfp$ the particle 
mean free path and $\nui$ the (ion) collision frequency. 
This requirement is rarely satisfied in real turbulent astrophysical and space 
plasmas (e.g., in the solar wind, $\mfp\sim$1~AU) and it is an observational 
certainty that turbulence exists at collisionless scales 
\cite{Bruno_Carbone,Bale_etal}. 
The same is true in fusion plasmas. Thus, plasma turbulence must be understood 
in the framework of kinetic theory, which evolves the distribution function $f_s$ for each 
species~$s$~($=i,e$): 
\bea
\label{Vlasov}
{\dd f_s\over\dd t} + \vv\cdot\vdel f_s + 
{\qs\over m_s}\lt(\vE + {\vv\times\vB\over c}\rt)\cdot{\dd f_s\over\dd\vv} 
= \dtcolls,
\eea
where $\qs$ and $m_s$ are particle charge and mass, 
$c$ is the speed of light, the right-hand side of \eqref{Vlasov} 
is the collision integral (quadratic in $f$), 
and $\vE$ and $\vB$ are the electric and magnetic fields, which satisfy 
Maxwell's equations: 
\bea
\label{Poisson}
\vdel\cdot\vE = 4\pi\sum_s\qs n_s,\qquad\qquad\qquad\ \, n_s = \intv f_s,\\
\label{Ampere}
\vdel\times\vB - {1\over c}{\dd\vE\over\dd t} = {4\pi\over c}\lt(\vj+\vja\rt),\qquad
\vj = \sum_s\qs\intv\vv f_s,\\
\label{Faraday}
{\dd\vB\over\dd t} = -c\vdel\times\vE,\qquad \vdel\cdot\vB=0.
\eea
In \eqref{Ampere}, the external current 
$\vja$ stands in for the outer-scale energy injection. 

The energy injected into the plasma must be dissipated and converted 
into particle heat. It is in fact a rather subtle issue what this exactly means. 
Multiplying \eqref{Vlasov} by $m_sv^2/2$ and integrating, we 
find that the total particle energy satisfies:
\bea
\fl
{\rmd\over\rmd t}\intrV\sum_s\intv{m_s v^2\over2}\,f_s 
%= \intrV\sum_s\qs\intv\vE\cdot\vv 
= \intrV\vE\cdot\vj
= \eps - {\rmd\over\rmd t}\intrV {E^2 + B^2\over8\pi},
\label{work_done}
\eea 
where $\eps=-(1/V)\intr\vE\cdot\vja$ is the injected power per unit volume. 
In deriving the above equation, we used Amp\`ere's law \eqref{Ampere}, 
Faraday's law \eqref{Faraday}, and integrated by parts wherever opportune.
\Eqref{work_done} tells us that, unsurprisingly, the change in particle energy is 
equal to the work done on the particles ($\intr\vE\cdot\vj$) and that 
the change in the combined energy of the particles and fields is equal 
to the injected energy. This, however, is not yet a statement about heating 
in the thermodynamic sense of the term because the energy exchange described 
by \eqref{work_done} is, in principle, reversible. In order to effect irreversible 
heating, we must change the entropy of the system and that, in a closed kinetic system, 
can only be accomplished by collisions. This result is known as Boltzmann's $H$-theorem 
\cite{Boltzmann}: from \eqref{Vlasov}, it is readily obtained \cite{Longmire} 
that the entropy $S_s$ of species $s$ grows according 
to\footnote{Boltzmann's function is $H=(1/V)\intr\intv f\ln f
=-S$, so $\rmd H/\rmd t\le0$.} 
\bea
\fl
{\rmd S_s\over\rmd t} \equiv {d\over dt}\lt[-\intrV\intv f_s\ln f_s\rt] 
= -\intrV\intv\ln f_s\dtcolls \ge 0.
\label{Boltzmann}
\eea

We would now like to assume that the 
plasma distribution function can be split into a slowly changing equilibrium 
part and a fast changing fluctuating part, $f_s=\fMs + \dfs$, that 
the latter is small, and that its smallness is controlled by some 
parameter $\epsilon\ll1$. In the next section, we shall specialize to the 
case of gyrokinetic turbulence, where $\epsilon\sim\omega/\Omega_i$, 
the ratio of the typical fluctuation frequency to the ion cyclotron frequency. 
As we shall see momentarily, the equilibrium quantities can be 
assumed to vary on a time scale $\sim(\epsilon^2\omega)^{-1}$, much longer 
than the fluctuation time scale $\omega^{-1}$. 
We further assume that the collision rate is 
$\nui\sim\omega$, i.e., while the dynamics are not collisionally 
dominated, collisions are retained on a par with fluctuations. 
This can be viewed as a convenient ordering prescription 
on the level of the $\epsilon$ expansion \cite{Howes_etal} 
and does not prevent one from considering the collisional 
($\nui\gg\omega$) and collisionless ($\nui\ll\omega$) regimes as subsidiary 
limits \cite{Tome}. With these assumptions, \eqref{Boltzmann} implies 
that the equilibrium distribution is a local Maxwellian 
for each species \cite{Boltzmann,Longmire}: 
$\fMs = {\ns}(\pi\vths^2)^{-3/2}\exp(-v^2/\vths^2)$, 
where $\vths=(2\Ts/m_s)^{1/2}$ is the thermal speed and $\Ts$ the temperature. 
For simplicity, we shall ignore all spatial gradients of the equilibrium 
quantities compared to the gradients of the fluctuating ones 
and also assume that the plasma motions are subsonic, 
i.e., the Mach number is small, $M=u/\vths\sim\epsilon\ll1$.\footnote{These are
rarely good assumptions at the outer scale, but, in many astrophysical 
applications, they are increasingly better satisfied
as we move deeper into the inertial range \cite{Tome}. 
In tokamak plasmas, the equilibrium gradients do play an important role, 
but it is not essential to retain them in the conceptual 
discussion that follows.} 

If we now substitute $f_s=\fMs + \dfs$ into \eqref{Boltzmann}, 
use the assumptions explained above, and keep only the lowest-order 
terms in $\epsilon$, we~get
\bea
\fl
\nonumber
\Ts\,{\rmd S_s\over\rmd t} &=& 
{\rmd\over\rmd t}\lt[\intrV\intv{m_s v^2\over2}\,f_s
-\intrV\intv{\Ts\dfs^2\over2\fMs}\rt]\\
\fl
&=& -\intrV\intv{\Ts\dfs\over\fMs}\ddtcolls + \intv{m_s v^2\over2}\dMtcolls.
\label{entropy}
\eea 
The second term on the right-hand side represents the collisional 
energy exchange between the Maxwellian equilibria of two species 
and is equal to $-\ns\nu_E^{ss'}(\Ts-\Tsp)$, where 
$\nu_E^{ss'}$ is the appropriate rate of collisions 
between species $s$ and $s'$ \cite{Helander_Sigmar}. 
\Eqref{entropy} has two key consequences. First, let us average it over 
times longer than the fluctuation time scale but shorter than the 
equilibrium-variation time scale, $\omega^{-1}\ll t\ll(\epsilon^2\omega)^{-1}$. 
Then the time derivatives of the fluctuating quantities vanish and, 
noting that $\intv(m_sv^2/2)\fMs = (3/2)\ns\Ts$ and $\rmd\ns/\rmd t=0$, 
we get \cite{Howes_etal}
\beq
\fl
{3\over2}\,\ns{\rmd\Ts\over\rmd t} = -\overline{\intrV\intv{\Ts\dfs\over\fMs}\ddtcolls} 
-\ns\nu_E^{ss'}(\Ts-\Tsp),
\label{heating}
\eeq
where the overline denotes the time average. The first term on the right-hand side
is positive definite and represents the heating of the equilibrium via collisional 
dissipation of the fluctuating part of the distribution function --- precisely 
the transfer of the fluctuation energy into heat that is the ultimate 
imperative of turbulence. Note that \eqref{heating} is consistent with 
the ordering assumptions made earlier: the equilibrium evolves 
on the time scale $\sim(\epsilon^2\omega)^{-1}$, as we have ordered $\nuss\sim\omega$. 

The second important consequence of \eqref{entropy} arises 
if we sum over species and use \eqref{work_done} to express the first term 
under the time derivative in \eqref{entropy}. This gives 
\beq
\fl
%{\rmd W\over\rmd t} \equiv
{\rmd\over\rmd t}\intrV\lt[\sum_s\intv{\Ts\dfs^2\over2\fMs} + {E^2+B^2\over8\pi}\rt] 
= \eps + \intrV\sum_s\intv{\Ts\dfs\over\fMs}\ddtcolls.
\label{W_cons}
\eeq
The positive definite quantity 
under the time derivative on the left-hand side, henceforth denoted $W$, 
will be referred to as {\em generalized energy}.\footnote{We use this term to emphasize 
the role of $W$ as the cascaded quantity in plasma turbulence (see below). 
The importance of its conservation for plasma turbulence was realized 
by several authors \cite{Fowler,Hallatschek,Watanabe_Sugama,Howes_etal,Scott}, 
who refer to it as the ``generalized grand canonical potential'' or free energy.
The latter term is perhaps physically the most appropriate because it 
flags the interpretation of $W$ as the work 
content of the particles $+$ fields system. 
The part of $W$ that involves $\dfs$ is equal 
to $-\sum_s\Ts\delta S_s$, where $\delta S_s$ is the perturbed 
entropy. In an exactly collisionless plasma, \eqref{entropy} 
and \eqref{work_done} show that any work done on the plasma 
simply increases this quantity. Any increase in the ``equilibrium'' entropy 
that might appear to be heating is then, in fact, compensated by a decrease 
in the perturbed entropy, so this ``heating'' is reversible. 
See \cite{Krommes_Hu,Krommes_df,Sugama_etal} for discussion of the entropy production 
in plasmas.} 
Its evolution is determined by the competition (or, in a stationary state, balance) of the 
externally supplied power $\eps$ and collisional dissipation (the negative-definite 
term on the right-hand side) --- the latter converts the generalized energy into 
heat according to \eqref{heating}. Thus, we have a conservation law analogous to 
\eqref{NS_energy}. This suggests a straightforward generalization of the view of fluid 
turbulence outlined in \secref{sec_intro} to plasma turbulence: its cause and effect
is the transfer of the generalized energy injected at the outer scale 
to scales where the collisional dissipation can convert it to heat. 

There is, however, an important novel feature here. 
If the collision frequency is small, $\nuss\ll\omega$,  
the collision term in \eqref{W_cons} can only balance the injected power provided 
the perturbed distribution function develops small-scale structure in 
velocity space. Since the collision operator is a second-order (diffusion) 
operator in the velocity space, we may roughly estimate 
the smallness of this structure by balancing $\omega\sim\nuss\vths^2\dd^2/\dd v^2$, 
so the correlation scale in velocity space is 
$\delta v/\vths\sim (\nuss/\omega)^{1/2}$. As we shall see in \secref{sec_ent}, 
in gyrokinetic turbulence, the emergence of small scales in velocity space 
is intertwined with a cascade to small scales in physical space. 
Thus, in the same way as fluid turbulence could be described as 
the energy cascade, plasma turbulence is a cascade of generalized energy, 
or a {\em kinetic cascade} --- this cascade occurs in phase space, reaching 
towards small scales both in physical space and in velocity space. 

If the heating is always ultimately collisional, what then is the 
status of the collisionless (Landau) damping \cite{Landau_damping} 
as a dissipation mechanism for 
(homogeneous) plasma turbulence? Collisionless damping does not appear explicitly in \eqref{W_cons} 
because what it does is, in fact, redistribute the generalized 
energy: the energy of electromagnetic fluctuations ($E^2+B^2$) is converted 
into entropy fluctuations ($\Ts\dfs^2/2\fMs$). In order for any 
actual heating to occur (i.e., for the fluctuation energy to be lost 
irreversibly), this perturbed entropy has to be transferred through phase space to 
collisional scales. There are two ways in which this can be accomplished: 
linear and nonlinear. The first is the well known \cite{Hammett_Dorland_Perkins,Stix} 
phase-mixing mechanism associated with the so-called ballistic 
response in the perturbed distribution function: the linearized kinetic 
equation \eqref{Vlasov} has the homogeneous solution  
$\dfs \propto \rme^{-\rmi\vk\cdot\vv t}$ \cite{Landau_damping}, for which  
$\dd\dfs/\dd v\sim kt\,\dfs$, i.e., there is a secular 
growth of the velocity-space derivatives and the collisions become important 
after a time $t\sim (k\vths)^{-1}(\omega/\nuss)^{1/2}$. 
In fact, as anticipated in \cite{Dorland_Hammett} and 
as we will show in \secref{sec_ent}, the linear phase 
mixing can be superceded by a faster nonlinear mechanism that 
cascades the perturbed entropy to collisional velocity scales 
over times~$t\sim\omega^{-1}$. 

Finally, we note that one can make a plausible argument in favour of an effectively 
irreversible ``collisionless heating'' in the sense that the distribution function 
may become so convoluted in phase space that it is effectively impossible to unscramble 
it and the entropy of an approppriately defined ``coarse-grained'' distribution 
is increased. Discussions of this process and the difference between such effective 
irreversibility and the exact irreversibility for which collisions are 
necessary have continued since the birth of quasilinear theory to 
the present day.\footnote{Understanding the role of phase mixing and weak collisions 
in converting wave energy into heat is practically important, e.g., in the theory of RF heating; 
see, e.g., \cite{Stix,Bilato_Brambilla} and references therein.} 
Here we only need to emphasize 
the salient physical fact that until the collsions can act, the negative 
entropy necessary to compensate for the increase in the coarse-grained entropy 
is stored in the fluctuations of the perturbed distribution function and that 
these fluctuations are explicitly present in the overall generalized 
energy budget \eqref{W_cons} --- a conservation law that underpins the 
interpretation of plasma turbulence proposed here. 
 
\section{Gyrokinetics and the Many Forms of the Kinetic Cascade}
\label{sec_GK}

Our treatment so far has not been specific to gyrokinetic turbulence. 
However, the particular mechanism of kinetic cascade in phase space 
we intend to discuss in \secref{sec_ent} will be. Thus, we now briefly 
introduce the gyrokinetic approximation and describe the forms the kinetic 
cascade from macro to microscales takes in gyrokinetic turbulence. 

It is nature's gift to plasma physicists that 
magnetized plasma turbulence both in fusion devices and in space appears 
to consist mostly of fluctuations whose frequencies are much lower than the 
ion cyclotron frequency, $\omega\ll\Omega_i$, even as their spatial 
scales perpendicular to the magnetic field 
can be as small as or smaller than the ion gyroscale $\rho_i=\vthi/\Omega_i$. 
This low-frequency character of the turbulence is intimately related 
to the tendency of plasma fluctuations in a dynamically 
strong magnetic field to be spatially anisotropic, with $\kpar\ll\kperp$. 
Let us briefly explain why. 

The structure of plasma turbulence is set by  
the interplay of parallel linear propagation effects (waves, particle streaming) 
and perpendicular nonlinear decorrelation (turbulent cascade). 
It is crucial to understand that, while anisotropic, this is an essentially 
three-dimensional situation. For fluctuations with a given perpendicular 
correlation length, the parallel correlation length 
is set by the distance a wave (or streaming particles) can travel 
during one perpendicular correlation time.\footnote{Clearly, perpendicular 
planes separated by longer distances cannot remain correlated, which rules out  
the two-dimensional limit (linear frequency $\ll$ 
nonlinear decorrelation rate). Decorrelation at shorter distances gives 
rise to weak turbulence (linear frequency $\gg$ nonlinear decorrelation rate), 
which tends to produce a cascade towards smaller perpendicular scales, where 
the nonlinear decorrelation rate again becomes comparable to 
the wave frequency \cite{GS97,Galtier_etal}.} 
A good example of this principle is the Alfv\'enic MHD turbulence, 
where it is known as the {\em critical balance} \cite{GS95,GS97}. 
Alfv\'enic turbulence is the predominant type of turbulence 
in finite-beta plasmas at scales above the ion gyroscale 
(the ``inertial range'') irrespective of the degree of collisionality
--- this statement can be proven analytically \cite{SCD_kiev,Tome} 
and there is ample evidence in its favour from measurements 
in the solar wind \cite{Bruno_Carbone,Bale_etal}. Alfv\'enic fluctuations have velocities 
and perturbed magnetic fields $\uperp\sim\dBperp/\sqrt{4\pi m_i\ni}$ 
perpendicular to the mean field $\vB_0 = B_0\vz$. 
Their decorrelation rate is $\sim\kperp\uperp$, while 
the characteristic propagation frequency is 
$\omega=\kpar v_A$, where $v_A=B_0/\sqrt{4\pi m_i\ni}$. 
In critical balance, $\kpar v_A\sim\kperp\uperp$, so 
$\kpar/\kperp\sim\uperp/v_A\ll1$. If the Alfv\'enic cascade 
from the outer scale to the ion gyroscale respects this 
principle\footnote{{\em A one-paragraph review of the Goldreich--Sridhar 1995 
MHD turbulence theory} \cite{GS95,GS97}: Making the same assumptions as in 
Kolmogorov's theory (footnote \ref{fn_K41}) except isotropy, we have, for Alfv\'enic 
velocities, $\ul^2/\taul\sim\eps$, where $\lambda$ is now the perpendicular scale. 
If the critical balance holds, 
$v_A/\lparl\sim\ul/\lambda$, where $\lparl$ is the parallel correlation length 
of these fluctuations. Since this means that only one time scale is present in the problem, 
we must have $\taul\sim\lambda/\ul$ and thus recover the Kolmogorov 
scaling: $\ul\sim(\eps\lambda)^{1/3}$. Using this and the critical balance, we find the scaling 
relationship between the perpendicular and parallel scales: 
$\lparl\sim\leps^{1/3}\lambda^{2/3}$, where $\leps = v_A^3/\eps$
(but see \cite{Boldyrev} for a version of this theory giving rise to different scalings). 
Thus, there is a cascade both in the parallel and perpendicular directions, but the 
aspect ratio $\lparl/\lambda$ increases as we move deeper into the inertial~range.\label{fn_GS}} 
(and there is numerical \cite{CV_aniso,Maron_Goldreich} 
and observational \cite{Horbury_etal_aniso} evidence that it does), 
the fluctuation frequency at $\kperp\rho_i\sim1$ will still be 
low compared to the ion cyclotron frequency: 
$\omega/\Omega_i\sim\kpar v_A/\Omega_i\sim (\kpar/\kperp)\kperp\rho_i/\sqrt{\beta_i}\ll1$
(we assume moderate values of $\beta_i$). 

The gyrokinetic approximation can now be constructed by 
using the critical balance explicitly as the ordering prescription: 
$\epsilon\sim\kpar/\kperp\sim\omega/\Omega_i\sim\uperp/v_A\sim \qs\ephi/\Ts 
\sim\dBperp/B_0\sim\dBpar/B_0\sim \dfs/\fMs$, where $\ephi$ is 
the scalar potential.\footnote{The Alfv\'enic velocity perturbation 
is the $\vE\times\vB_0$ flow: $\vuperp= \vz\times\vdperp c\ephi/B_0$.} 
The Vlasov--Maxwell equations \eqsdash{Vlasov}{Faraday} are expanded in $\epsilon$ 
and averaged over the particle gyromotion \cite{Frieman_Chen,Howes_etal,Brizard_Hahm}. 
As a result of this procedure, the perturbed distribution function splits 
into the Boltzmann response and the perturbed distribution of particle 
gyrocentres: $\dfs = -\qs\ephi\fMs/\Ts + \hs(t,\vR_s,\vperp,\vpar)$, 
where $\vR_s=\vr + \vvperp\times\vz/\Omega_s$ is the gyrocentre position. 
In a uniform magnetic field $\vB_0$, the gyrokinetic equation for $\hs$ is 
\bea
\label{GK}
{\dd\hs\over\dd t} + \vpar\,{\dd\hs\over\dd z} + 
{c\over B_0}\lt\{\avchi,\hs\rt\} = {\qs\fMs\over\Ts}{\dd\avchi\over\dd t} 
+ \dhtcolls,
\eea
where $\chi = \ephi - \vv\cdot\vA/c$, $\vB= B_0\vz+\dvB$, $\dvB=\vdel\times\vA$, 
$\vdel\cdot\vA=0$, and $\la\cdots\ra_{\vR_s}$ is the gyroangle average at constant $\vR_s$. 
The vector potential $\vA$ is recovered from \eqref{Ampere} neglecting the displacement current. 
The scalar potential $\ephi$ is found from the quasineutrality condition: 
neglecting $\vdel\cdot\vE$ in \eqref{Poisson} and separating the Boltzmann 
response, we have 
\bea
\sum_s {\qs^2\ephi\over\Ts}\,\ns = \sum_s \qs\intv \la\hs\ra_\vr,
\label{quasineut}
\eea
where $\la\cdots\ra_\vr$ denotes gyroaveraging at constant $\vr$
(the velocity integral is at constant~$\vr$). 

Gyrokinetics helps make the problem of kinetic cascade 
numerically \cite{Howes_etal3,Tatsuno_etal} and, in certain 
limits, analytically \cite{Tome,Plunk_PhD} tractable 
because all high-frequency physics ($\omega\ge\Omega_i$) 
is systematically ordered out
and the gyroaveraging reduces the phase space from 6D to 5D. 
However, it is still a fully kinetic system and everything that was 
said about heating and the kinetic cascade in \secref{sec_gen} 
remains valid. The generalized energy conservation law \eqref{W_cons} 
for gyrokinetics takes the following form \cite{Howes_etal,Tome}:
\bea
\nonumber
{\rmd W\over\rmd t} &=& 
{\rmd\over\rmd t}\intrV\lt[\sum_s\lt(\intv{\Ts\la\hs^2\ra_\vr\over2\fMs}
- {\qs^2\ephi^2\ns\over2\Ts}\rt) + {|\dvB|^2\over8\pi}\rt]\\
&=& \eps + \sum_s\intv\intRs{\Ts\hs\over\fMs}\dhtcolls.
\label{W_gk}
\eea
As we explained in \secref{sec_gen}, the generalized energy injected 
at the outer scale has to be transferred (cascaded) through phase space eventually 
to reach the collisional scales. If we conjecture that this transfer 
is local in scale space, 
we can ask what forms the kinetic cascade takes in several distinct 
physical regimes separated by the characteristic plasma microscales: 
the mean free path, the ion and the electron gyroscales. 
It turns out that in each of the asymptotic limits 
$\kpar\mfp\ll1$, $\kpar\mfp\gg1$, $\kperp\rho_i\ll1$, 
$\kperp\rho_i\gg1$, etc., the kinetic cascade splits into several
non-energy-exchanging channels corresponding to cascades of distinct 
plasma fluctuation modes, some of which are familiar from 
fluid models of plasma turbulence and some are new. 
As the characteristic scales are crossed ($\kpar\mfp\sim1$, $\kperp\rho_i\sim1$), 
these channels join together into a single cascade, which 
then splits again, but in a different way, as another asymptotic 
limit is reached. 
All these asymptotic limits are worked out in detail in \cite{Tome}. 
Here we briefly summarize their role as a route 
for the generalized energy to reach the ion gyroscale (at which point 
interesting things start happening in the phase space). 

Let us imagine that energy is injected at scales larger 
than both the mean free path and the ion gyroradius. As the cascade takes the 
energy to smaller scales, anisotropy and critical balance are established, so 
the gyrokinetic approximation applies \cite{Howes_etal2,Howes_DPP}. 
In the inertial range ($\kperp\rho_i\ll1$), the energy cascade is split into 
two main channels: the Alfv\'enic turbulence ($\dvBperp$, $\vuperp$), which is described by the 
Reduced MHD equations \cite{Strauss} regardless of the collisionality \cite{SCD_kiev,Tome} 
and the ``compressive'' component ($\delta n$, $\upar$, $\dBpar$). 
The Alfv\'enic cascade is split into two cascades corresponding 
to the two directions of propagation of the Alfv\'en waves. 
While the nonlinear interaction is always between the ``$+$'' and ``$-$'' waves, 
it is of ``scatter'' type, so no energy is exchanged between the two cascades.  
The compressive fluctuations are passively mixed by the Alfv\'en waves, again 
without energy exchange. In the collisional limit ($\kpar\mfp\ll1$), the 
compressive cascade is split into three channels: the ``$+$'' and ``$-$'' slow waves and the entropy-mode. 
As $\kpar\mfp\sim1$ is approached, these three are mixed together and remain mixed 
for $\kpar\mfp\gg1$. Dissipation and collisional heating can occur at this 
transition because in the fluid limit, the collisional term in \eqref{W_cons} can be 
activated by small deviations of the distribution function from a Maxwellian 
--- the smallness of the collision rate in this case is overcome not by a 
velocity-space cascade but by the fact that the non-Maxwellian part of the perturbed 
distribution function is proportional to $\kpar/\nui$. At collisionless scales, 
the compressive fluctuations experience the Barnes (transit-time) \cite{Barnes,Stix} 
version of Landau damping --- as discussed in \secref{sec_gen}, this transfers 
the energy associated with the compressive fluctuations into ion entropy fluctuations. 

As the inertial-range cascade transfers energy to scales around $\kperp\rho_i\sim1$, 
its Alfv\'enic and compressive components cease to be decoupled from each other 
and all fluctuations 
are subject to Landau damping (see \cite{Howes_etal} for details of linear gyrokinetics). 
What emerges on the other side of this transition, 
at $\kperp\rho_i\gg1$,\footnote{In space physics, this is called the ``dissipation range,''
a historical misnomer dating back to the times when it was not appreciated that it can contain 
dissipationless cascades.} is a cascade of generalized energy again split into two channels: 
the fluctuations polarized as kinetic Alfv\'en waves (KAW) (they satisfy fluid-like 
equations closely related to Electron MHD \cite{Kingsep_Chukbar_Yankov,Tome}) 
and, energetically decoupled from them, the ion entropy fluctuations ($\Ti\hi^2/2\fMi$).
The latter carry the part of the inertial-range energy that was Landau-damped 
at the ion gyroscale and, possibly, also in the inertial range (for the compressive 
fluctuations). How it becomes ion heating is the subject of \secref{sec_ent}.
The KAW cascade takes the energy to the electron gyroscale, $\kperp\rho_e\sim1$, 
where it is converted by Landau damping into electron entropy fluctuations
($\Te\he^2/2\fMe$), eventually giving rise to electron heating in a way analogous 
to the ion case discussed below. 

\section{Nonlinear Perpendicular Phase Mixing and the Entropy Cascade}
\label{sec_ent}

In order to introduce the concept of the phase-space cascade of entropy in the simplest 
possible setting, we will consider the extreme case where all of the fluctuation 
energy arriving to the ion gyroscale from the inertial range is converted into 
entropy fluctuations by Landau damping, i.e., we will neglect the KAW component 
of the dissipation-range turbulence.\footnote{In the presence of KAW turbulence, 
the entropy fluctuations are passively mixed by KAW. This case can be treated 
in a way analogous to what we do here \cite{Tome}. At moderate values of 
$\beta_i$, a KAW cascade is probably a good description of the dissipation-range 
turbulence in the solar wind \cite{Howes_etal2,Howes_etal3}. The case without KAW may be 
more relevant at low and high $\beta_i$ because ion Landau damping is quite strong 
then \cite{Howes_etal}. It is also interesting in the context of 
electrostatic microturbulence (ITG, ETG, drift waves) 
prevalent in fusion plasmas \cite{Cowley_Kulsrud_Sudan,Dorland_ETG,Horton}.} 
Furthermore, we will use the Boltzmann-electrons approximation, which is justified 
to the lowest order in the mass-ratio expansion as long as we stay 
above the electron gyroscale, $\kperp\rho_e\ll1$ \cite{Snyder_Hammett,Tome}. 
These approximations mean that we have $\dfe=e\ephi\fMe/\Te$, i.e., $\he=0$, 
while $\hi$ satisfies the electrostatic version of \eqref{GK} ($\chi=\ephi$). 
The resulting system of equations follows from \eqsand{GK}{quasineut}: 
\bea
\label{GK_ions}
\fl
{\dd\hi\over\dd t} + \vpar\,{\dd\hi\over\dd z} + 
{c\over B_0}\lt\{\avphi,\hi\rt\} -\dhtcolli = {\dd\over\dd t}{Ze\avphi\over\Ti}\,\fMi,\\
\label{quasineut_ions}
\fl
\lt(1+{\tau\over Z}\rt){Ze\ephi\over\Ti} = {1\over\ni}\intv \la\hi\ra_\vr 
= \sum_\vk\rme^{\rmi\vk\cdot\vr}{1\over\ni}\intv J_0\lt({\kperp\vperp\over\Omega_i}\rt)\hik,
\eea
where $Z=\qi/e$, $\tau=\Ti/\Te$ %$J_0$ is a Bessel function 
and $\hik$ is the Fourier transform of $\hi(\vR_i)$.

As we explained in \secref{sec_gen}, in order for the collision term in 
\eqref{GK_ions} to become non-negligible, small-scale structure has to be 
generated in the velocity space with $\delta v/\vthi\sim(\nui/\omega)^{1/2}$. 
One route to such small scales is via the parallel (linear) phase mixing, whose role 
in plasma turbulence has been well established for some time 
\cite{Hammett_Dorland_Perkins,Krommes_Hu,Krommes_df,Watanabe_Sugama}: 
the ballistic response $\hi\propto\rme^{\rmi\kpar\vpar t}$ gives rise to 
secularly growing gradients $\dd\hi/\dd\vpar\sim\kpar t\,\hi$ and, therefore, 
small scales in parallel velocities: $\delta\vpar\sim {1/\kpar t}$.

\begin{figure}[t]
\centerline{\epsfig{file=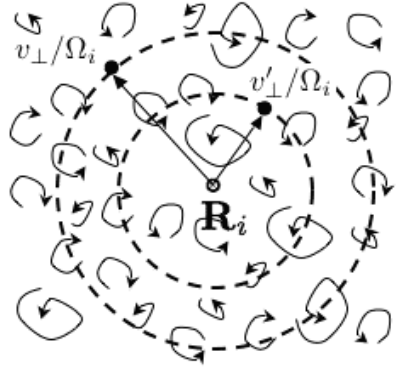,width=3.5in}}
\caption{\label{fig_gyroorbits}The nonlinear perpendicular phase 
mixing: the gyrocentre distribution function at a given point $\vR_i$
is mixed in a decorrelated way by $\vE\times\vB$ flows gyroaveraged 
over ion orbits whose radii ($\vperp/\Omega_i$ and $\vperp'/\Omega_i$) 
differ by more than the flows' correlation~length.}
\end{figure}

The other, perpendicular, phase mixing mechanism is nonlinear \cite{Dorland_Hammett,Tome}. 
In \eqref{GK_ions}, the nonlinear 
term represents mixing of the ion distribution in the gyrocentre space 
by the gyroaveraged $\vE\times\vB$ flows. Like any random mixing, this produces 
small scales in $\vR_i$. It also produces small scales in $\vperp$ for the following 
reason. Consider \eqref{GK_ions} taken at two different values of velocity, 
$\vperp$ and $\vperp'$. Then $\hi(\vR_i,\vperp)$ and $\hi(\vR_i,\vperp')$ 
will be spatially mixed by the gyroaveraged $\vE\times\vB$ velocity field given by 
$\avphi(\vperp)$ and $\avphi(\vperp')$, respectively. These gyroaverages come from spatially 
decorrelated fluctuations of $\ephi$ if the difference between the 
gyroradii $\vperp/\Omega_i$ and $\vperp'/\Omega_i$ is larger than the perpendicular correlation 
length $1/\kperp$ of $\ephi$ (\figref{fig_gyroorbits}). 
If this condition is satisfied, $\hi(\vR_i,\vperp)$ 
and $\hi(\vR_i,\vperp')$ are mixed by decorrelated fields and are, therefore, 
themselves decorrelated. Thus, small-scale structure of $\ephi$ in 
the physical space gives rise to small-scale structure of $\hi$ in the velocity space:\footnote{Note 
that this nonlinear perpendicular phase mixing mechanism was first recognized in 
\cite{Dorland_Hammett}: in their gyrofluid closure formalism, it manifested itself 
as the growth of high-order $\vperp$ moments of $\hi$.} 
the correlation scale in the velocity space is  
\bea
{\delta\vperp\over\vthi} = {1\over\rho_i} \lt|{\vperp\over\Omega_i}-{\vperp'\over\Omega_i}\rt| 
\sim {1\over\kperp\rho_i} \ll1\quad \mathrm{when}\quad \kperp\rho_i\gg1.
\label{vperp_scale}
\eea 

The small-scale structure of $\hi$ in the gyrocenter 
space gives rise to similarly small-scale structure of $\ephi$ in the physical space. 
Using \eqref{quasineut_ions}, they can be related as follows. 
For $\kperp\rho_i\gg1$, the Bessel function in the velocity integral is 
$J_0(\kperp\vperp/\Omega_i)\simeq (2\Omega_i/\pi\kperp\vperp)^{1/2}\cos(\kperp\vperp/\Omega_i-\pi/4)$, 
i.e., it oscillates in $\vperp$ with the period 
$\delta\vperp/\vthi=2\pi/\kperp\rho_i$. 
But, according to \eqref{vperp_scale}, this is also the 
correlation scale of $\hik$ in velocity space. Assuming that the velocity 
integral accumulates as a random walk and taking into account also 
the factor of $1/\sqrt{\kperp\rho_i}$ coming from the Bessel function, we have 
\bea
{Ze\phik\over\Ti}\sim {\vthi^3\over\ni}{1\over\sqrt{\kperp\rho_i}}
\lt({\delta\vperp\over\vthi}\rt)^{1/2}\hik\sim {\vthi^3\over\ni}{\hik\over\kperp\rho_i}.
\label{phi_vs_hi}
\eea
The gyroaveraged potential is then 
$Ze\avphik/\Ti = Ze J_0(\kperp\vperp/\Omega_i)\phik/\Ti \sim (\vthi^3/\ni)\hik/(\kperp\rho_i)^{3/2}$, 
and so the perpendicular mixing of the particle distribution is a fully 
nonlinear process. 

This process can be understood as a local (in scale) ion {\em entropy cascade} and 
a Kolmogorov-style scaling theory can be constructed for it. 
Recall that the gyrokinetic equation \eqref{GK_ions} has a conservation law 
given by \eqref{W_gk} with $\he=0$ and $\dvB=0$. In view of \eqref{phi_vs_hi}, 
$Z^2e^2\ephi^2\ni/2\Ti\ll\intv\Ti\hi^2/2\fMi$, so the entropy of the perturbed 
ion distribution is conserved individually. This is, in fact, 
obvious also from \eqref{GK_ions}: again using \eqref{phi_vs_hi}, 
the inhomogeneous term on the right-hand side is negligible for $\kperp\rho_i\gg1$
and $\int d^3\vR_i\,\hi^2$ is clearly a conserved quantity but for collisions. 
Denoting by $\phil$ and $\hil$ the characteristic fluctuation amplitudes 
at some perpendicular scale $\lambda\ll\rho_i$ and by 
$\taul$ the corresponding cascade time, we may write 
(cf.\ footnotes \ref{fn_K41} and \ref{fn_GS})
\bea
\label{const_flux}
{m_i \vthi^8\over\ni}{\hil^2\over\taul}\sim\eps,\qquad
\taul\sim \lt({\rho_i\over\lambda}\rt)^{1/2}{\lambda^2\over c\phil/B_0}
\sim {\rho_i^{1/2}\lambda^{1/2}\ni\over\vthi^4\hil},
\eea
where we used \eqref{phi_vs_hi} to get $c\phil/B_0\sim\vthi^4\hil\lambda/\ni$.
Combining these relations, we~find\footnote{It is also possible to 
derive exact scaling results analogous to Kolmogorov's 4/5 law, which prove to 
be consistent with \eqref{scalings} \cite{Plunk_PhD}.} 
\bea
\label{scalings}
\hil\sim {\ni\over\vthi^3} {\rho_i^{1/6}\lambda^{1/6}\over\leps^{1/3}},\qquad
{Ze\phil\over\Ti}\sim {\lambda^{7/6}\over\rho_i^{5/6}\leps^{1/3}},\qquad
\taul\sim {\leps^{1/3}\rho_i^{1/3}\lambda^{1/3}\over\vthi},
\eea
where $\leps = m_i\ni\vthi^3/\eps$. These scalings correspond 
to a $\kperp^{-4/3}$ spectrum of $\hi$ and a $\kperp^{-10/3}$ spectrum of $\ephi$. 
Encouragingly, these predictions seem to be corroborated by numerical simulations 
of electrostatic gyrokinetic turbulence in two spatial dimensions \cite{Tatsuno_etal}. 

Now let us revisit the question of parallel phase mixing. In our discussion of the 
perpendicular cascade so far, we have ignored the presence of the parallel propagation 
(particle streaming) term in \eqref{GK_ions}. In a formally 2D situation, i.e., 
when $\omega\sim\taul^{-1}\gg\kpar\vpar$, this is, of course, allowed and the 
perpendicular scalings derived above should hold. However, it is more likely 
that the parallel scale of the fluctuations will adjust to their perpendicular 
scale according to the critical balance principle explained in \secref{sec_GK}: 
$\lparl$ for the fluctuations with perpendicular scale $\lambda$ will be such 
that particles can stream across the distance $\lparl$ in one nonlinear decorrelation 
time $\taul$. Using \eqref{scalings}, 
this implies $\lparl\sim\vthi\taul\sim\leps^{1/3}\rho_i^{1/3}\lambda^{1/3}$
(cf.\ footnote \ref{fn_GS}). As we explained at the beginning of this section, 
the typical parallel correlation scale in velocity space produced by 
the parallel phase mixing is $\delta\vpar\sim1/\kpar t$. Therefore, after one 
perpendicular cascade time, no appreciable refinement of the parallel velocity-space 
structure is achieved: $\delta\vpar/\vthi\sim\lparl/\vthi\taul\sim1$. 
In contrast, in $\vperp$, one cascade time is enough for the entire cascade 
down to the collisional cutoff (to be calculated in \secref{sec_conc}) to be set up. 
Thus, the linear parallel phase mixing is much less efficient than the nonlinear 
perpendicular one. 
 
%Finally, let us consider another feature of the electrostatic gyrokinetics that we 
%have so far ignored: the presence of another invariant. 
%From \eqref{GK} with $\chi=\ephi$, it is not hard to show that the following 
%conservation law holds \cite{Plunk_PhD}\footnote{It is a particular consequence 
%of a larger set of conservation laws that in the most general case only 
%hold in 2D \cite{Plunk_PhD,Tome}.}
%\beq
%\fl
%-{\rmd\over\rmd t}\sum_s\sum_\vk {\qs^2\ns\over2\Ts}\lt[1-\Gamma_0(b_s)\rt]
%|\phik|^2 =\eps - \sum_s\qs\intv\intRs \avphis\dhtcolls,
%\eeq
%where $b_s=\kperp^2\rho_i^2/2$, $\Gamma_0(b_s)=I_0(b_s)\rme^{-b_s}$ and 
%$\eps = (1/V)\intr\Epar\jpar$. 

\section{Conclusion: Dissipation Achieved}
\label{sec_conc}

Let us now come back to the original motivation for the above developments: the necessity 
to understand how the distribution function is brought to collisional scales in the velocity space. 
We have seen that this is done by transferring the energy injected at the outer scale 
down to the ion and electron gyroscales via a multichannel cascade of generalized 
energy through phase space. Below the ion gyroscale, the phase-space nature of the 
cascade becomes particularly manifest as the ion distribution function simultaneously 
develops small scales in the gyrocentre and velocity space via a nonlinear perpendicular 
phase mixing process. We have described this process as a Kolmogorov-like turbulent 
cascade enabled by a constant flux of ion entropy and derived scaling relations for 
the fluctuations of the distribution function and the electric potential. 

Using these scalings \eqref{scalings}, let us now estimate the collisional cutoff 
in phase space. As we explained in \secref{sec_gen}, the collisional scale is reached if 
the velocity-space correlation scale is 
$\delta v/\vthi\sim(\nui/\omega)^{1/2}$. Using \eqref{vperp_scale} 
and estimating $\omega\sim\taul^{-1}$, we~get 
\bea
{\dvc\over\vthi}\sim {1\over\kc\rho_i} \sim \lt(\nui\taur\rt)^{3/5}\sim 
{\leps^{1/5}\rho_i^{2/5}\over\mfp^{3/5}},
\label{cutoff}
\eea 
where $\leps = m_i\ni\vthi^3/\eps$ and $\taur\sim(m_i\ni\rho_i^2/\eps)^{1/3}$ 
is the fluctuation time scale at $\kperp\rho_i\sim1$. 
This formula is perhaps our most consequential result for numerical applications: 
it tells us what it means to have a well-resolved gyrokinetic simulation of plasma 
turbulence and shows that the resolution requirements in the gyrocentre and velocity 
spaces are fundamentally linked. In this context, it is clear that adequate modeling 
of collisions \cite{Abel_etal,Barnes_etal} 
and controlled velocity-space resolution \cite{Barnes_vres} 
are imperative for gyrokinetic simulations. 

No matter how small the collisional cutoff \eqref{cutoff} is, all of the energy 
channelled into the sub-gyroscale entropy cascade will 
reach this cutoff in finite time --- roughly the nonlinear interaction scale $\taur$ 
evaluated at the ion gyroscale. Since the process is nonlinear, this time is 
amplitude dependent. If the principle of critical 
balance (\secref{sec_GK}) holds at the ion gyroscale, $\taur$ should be roughly equal 
to the linear parallel propagation time scale at $\kperp\rho_i\sim1$. 
Importantly, the time to reach the collisional cutoff does not depend on the 
collision rate --- just like in hydrodynamic turbulence (\secref{sec_intro}), 
the time to reach the viscous scale is the turnover time at the outer scale, 
independent of viscosity. 

Another important conclusion is that  
the dissipation range ($\kperp\rho_i>1$), even in the absence of kinetic Alfv\'en waves, 
is filled with electrostatic fluctuations due to the ion entropy cascade. This is a purely 
kinetic effect invisible in any fluid models. 
In fusion plasmas, this may be relevant for identifying the nature 
of electrostatic fluctuations found between the ion and electron gyroscales.\footnote{Another 
possible source of electrostatic fluctuations in the dissipation range is an inverse cascade 
of another conserved quantity of electrostatic gyrokinetics, $\intr\ephi^2$. 
Its conservation is a particular consequence of a larger set of more general 
gyrokinetic conservation laws valid in 2D (and only under some additional 
assumptions in 3D) \cite{Tome,Plunk_PhD}. However, since its cascade is inverse, 
it is only relevant if there is a source of energy below the ion gyroscale 
--- as, e.g., in the case of ETG turbulence~\cite{Dorland_ETG}.} 
In space physics, the great variability of the observed spectra in the 
dissipation range \cite{Smith_etal} might be speculatively attributed to 
varying proportions of energy contained in the entropy and 
kinetic-Alfv\'en-wave cascades~\cite{Tome}. 

These results are only the first glimpse of what one finds if one adopts 
the view of plasma turbulence as a kinetic cascade in phase space.
We believe that further studies conducted in this vein, 
both numerical \cite{Howes_etal3,Tatsuno_etal} and analytical \cite{Tome,Plunk_PhD}, 
will unveil much new physics and many new and tantalizing questions. 

\ack 
We gratefully acknowledge continued interactions with 
I~G~Abel, M~A~Barnes, C~H~K~Chen, T~S~Horbury, R~Numata and T~A~Yousef, 
who are involved in ongoing collaborations with us on some of the topics discussed in this paper. 
AAS was supported by an STFC Advanced Fellowship and by the STFC Grant ST/F002505/1. 
GGH and TT were supported by the US DOE Center for Multiscale Plasma Dynamics. 
WD, GWH, GGH, GGP and TT thank the Leverhulme Trust International Network for Magnetised 
Plasma Turbulence (Grant F/07 058/AP) for travel support.


\Bibliography{99}

\bibitem{Abel_etal}
Abel I G \etal 
%Barnes M A, Cowley S C, Dorland W and Schekochihin A A 
2008 \PP submitted

\bibitem{Bale_etal}
Bale S D \etal 
%Kellogg P J, Mozer F S, Horbury T S and Reme H 
2005 \PRL {\bf 94} 215002

\bibitem{Barnes}
Barnes A 1966 \PF {\bf 9} 1483

\bibitem{Barnes_vres}
Barnes M A 2008 {\it Ph.~D.\ Thesis} (University of Maryland) 

\bibitem{Barnes_etal}
Barnes M A \etal 
%Dorland W, Tatsuno T, Abel I G and Schekochihin A A 
2008 \PP submitted 

\bibitem{Bilato_Brambilla}
Bilato R and Brambilla M 2004 \PPCF {\bf 46} 1455

\bibitem{Boldyrev}
Boldyrev S A 2006 \PRL {\bf 96} 115002

\bibitem{Boltzmann}
Boltzmann L 1872 {\it Sitsungsber.\ Akad.\ Wiss.\ Wien}~{\bf 66} 275 
[English translation: in {\it Kinetic Theory~2} ed S G Brush (Oxford: Pergamon, 1966) 88]

\bibitem{Braginskii}
Braginskii S I 1965 {\it Rev.\ Plasma Phys.\ }{\bf 1} 205 

\bibitem{Brizard_Hahm}
Brizard A J and Hahm T S 2007 \RMP {\bf 79} 421

\bibitem{Bruno_Carbone}
Bruno R and Carbone V 2005 {\it Living Rev.\ Solar Phys.\ }{\bf 2} 4

\bibitem{CV_aniso}
Cho J and Vishniac E T 2000 \APJ {\bf 539} 273

\bibitem{Cowley_Kulsrud_Sudan}
Cowley S C, Kulsrud R M and Sudan R 1991 \PFB {\bf 3} 2767

%\bibitem{Diamond}
%Diamond P H \etal 2005 \PPCF {\bf 47} R35

\bibitem{Dorland_Hammett}
Dorland W and Hammett G W 1993 \PFB {\bf 5} 812

\bibitem{Dorland_ETG}
Dorland W \etal 2000 \PRL {\bf 85} 5579

\bibitem{Fowler}
Fowler T K 1968 {\it Adv.\ Plasma Phys.\ }{\bf 1} 201

\bibitem{Frieman_Chen}
Frieman E A and Chen L 1982 \PF {\bf 25} 502

\bibitem{Frisch}
Frisch U 1995 {\it Turbulence} (Cambridge: Cambridge University Press)

\bibitem{Galtier_etal}
Galtier S \etal, 
%Nazarenko S V, Newell A C and Pouquet A 
2000 {\it J.~Plasma Phys.\ }{\bf 63} 447

\bibitem{GS95}
Goldreich P and Sridhar S 1995 \APJ {\bf 438} 763

\bibitem{GS97}
Goldreich P and Sridhar S 1997 \APJ {\bf 485} 680

\bibitem{Hallatschek}
Hallatschek K 2004 \PRL {\bf 93} 125001

\bibitem{Hammett_Dorland_Perkins}
Hammett G W, Dorland W and Perkins F W 1991 \PFB {\bf 4} 2052

\bibitem{Helander_Sigmar}
Helander P and Sigmar D J 2002 {\it Collisional Transport in Magnetized Plasmas} 
(Cambridge: Cambridge University Press)

\bibitem{Horbury_etal_aniso}
Horbury T S, Forman M A and Oughton S 2008 \PRL submitted 
[\arXiv{0807.3713}]

\bibitem{Horton}
Horton W 1999 \RMP {\bf 71} 735

\bibitem{Howes_DPP}
Howes G G 2008 \PP {\bf 15} 055904

\bibitem{Howes_etal}
Howes G G \etal 
%Cowley S C, Dorland W, Hammet G W, Quataert E and Schekochihin A A 
2006 \APJ {\bf 651} 590

\bibitem{Howes_etal2}
Howes G G \etal
%Cowley S C, Dorland W, Hammet G W, Quataert E and Schekochihin A A 
2008 \JGR {\bf 113} A05103

\bibitem{Howes_etal3}
Howes G G \etal 
%Cowley S C, Dorland W, Hammet G W, Quataert E, Schekochihin A A and Tatsuno T 
2008 \PRL {\bf 100} 065004

%\bibitem{Jenko_Dorland}
%Jenko F and Dorland W 2002 \PRL {\bf 89} 225001

\bibitem{Kingsep_Chukbar_Yankov}
Kingsep A S, Chukbar K V and Yankov V V 1990 
{\it Rev.\ Plasma Phys.\ }{\bf 16} 243

\bibitem{K41}
Kolmogorov A N 1941 {\it Dokl.\ Akad.\ Nauk SSSR} {\bf 30} 299 
[English translation: {\it Proc\ Roy.\ Soc.\ London~A} {\bf 434} 9 (1991)]

\bibitem{Krommes_df}
Krommes J A 1999 \PP {\bf 6} 1477

\bibitem{Krommes_Hu}
Krommes J A and Hu G 1994 \PP {\bf 1} 3211

\bibitem{Landau_damping}
Landau L 1946 {\it J.~Phys.\ USSR} {\bf 10} 25 

\bibitem{Longmire}
Longmire C L 1963 {\em Elementary Plasma Physics} (New York: Interscience) 

\bibitem{Maron_Goldreich}
Maron J and Goldreich P 2001 \APJ {\bf 554} 1175

\bibitem{Obukhov}
Obukhov A M 1941 {\em Izv.\ Akad.\ Nauk SSSR Ser.\ Geogr.\ Geofiz.\ }{\bf 5} 453 

\bibitem{Plunk_PhD}
Plunk G G 2008 {\it Ph.~D.\ Thesis} (UCLA) 

\bibitem{Richardson}
Richardson L F 1922 {\it Weather Prediction by Numerical Process} 
(Cambridge: Cambridge University Press)

\bibitem{SCD_kiev}
Schekochihin A A, Cowley S C and Dorland W 2007 \PPCF {\bf 49} A195

\bibitem{Tome}
Schekochihin A A \etal 
%Cowley S C, Dorland W, Hammet G W, Howes G G, Quataert E and Tatsuno T 
2007 \APJS submitted [\arXiv{0704.0044}]

\bibitem{Scott}
Scott B D 2007 \PP submitted [\arXiv{0710.4899}]

\bibitem{Smith_etal}
Smith C W \etal 
%Hamilton K, Vasquez B J and Leamon R J 
2006 \APJ {\bf 645} L85

\bibitem{Snyder_Hammett}
Snyder P B and Hammett G W 2001 \PP {\bf 8} 3199

\bibitem{Stewart_movie}
Stewart R W 1969 {\it Turbulence} film 
directed by the National Committee for Fluid Mechanics Films, 
produced by EDC, Inc. 
[URL: http://web.mit.edu/fluids/www/Shapiro/ncfmf.html] 

\bibitem{Stix}
Stix T H 1992 {\it Waves in Plasmas} (New York: AIP)

\bibitem{Strauss}
Strauss H R 1976 \PF {\bf 19} 134

\bibitem{Sugama_etal}
Sugama H \etal
%Okamoto M, Horton W and Wakatani M 
1996 \PP {\bf 3} 2379

\bibitem{Tatsuno_etal}
Tatsuno T \etal 
%Barnes M A, Cowley S C, Dorland W and Schekochihin A A 
2008 \PRL submitted

\bibitem{Tsinober}
Tsinober A 2001 {\it An Informal Introduction to Turbulence} (Dordrecht: Kluwer)

\bibitem{Watanabe_Sugama}
Watanabe T-H and Sugama H 2004 \PP {\bf 11} 1476

\endbib
\end{document}